\documentclass[]{raa}            
\usepackage{graphicx,times}
\usepackage{amssymb}
\usepackage{amsmath}
\usepackage{natbib}
\begin{document}
   \title{An initial analysis of a strongly-lensed QSOs candidate identified by LAMOST
}

\volnopage{ {\bf 0000} Vol.\ {\bf 0} No. {\bf XX}, 000--000}
\setcounter{page}{1}

\author{Y. H. Chen\inst{1,2,3} \ M. Y. Tang\inst{1,2,4} \ H. Shu\inst{1,2,3} \ H. Tu\inst{5}}

\institute{\inst{1} Institute of Astrophysics, Chuxiong Normal University, Chuxiong 675000, China; {yanhuichen1987@126.com\\
           \inst{2} School of Physics and Electronical Science, Chuxiong Normal University, Chuxiong 675000, China\\
           \inst{3} Key Laboratory for the Structure and Evolution of Celestial Objects, Chinese Academy of Sciences, P.O. Box 110, Kunming 650011, China\\
           \inst{4} Department of Astronomy, Yunnan University, and Key Laboratory of Astroparticle Physics of Yunnan Province, Kunming, 650091, China\\
           \inst{5} Physics Department \& Shanghai Key Lab for Astrophysics, Shanghai Normal University, Shanghai 200234, China}
\\
\vs \no
{\small Received [0000] [July] [day]; accepted [0000] [month] [day] }}

\abstract{From 2011 to 2021, LAMOST has released a total of 76,167 quasar data. We try to search for gravitationally lensed QSOs by limiting coordinate differences and redshift differences of these QSOs. The name, brightness, spectrum, photometry and other information of each QSO will be visually checked carefully. Special attention should be paid to check whether there are groups of galaxies, gravitationally lensed arcs, Einstein crosses, or Einstein rings near the QSOs. Through careful selection, we select LAMOST J160603.01+290050.8 (A) and LAMOST J160602.81+290048.7 (B) as a candidate and perform an initial analysis. The component A and B are 3.36 arc seconds apart and they display blue during photometric observations. The redshift values of component A and B are 0.2\% different, their Gaia$\_$g values are 1.3\% different, and their ugriz values are 1.0\% or less different. For the spectra covering from 3,690 {\AA} to 9,100 {\AA}, the emission lines of C\,II, Mg, H\,$\gamma$, O\,III, and H\,$\beta$ are present for both component A and B and the ratio of flux(B) to flux(A) from LAMOST is basically a constant, around 2.2. However, no galaxies have been found between component A and B. Inada et al. identified them as binary quasars. But we accidentally find a galaxy group near the component A and B. If the center of dark matter in the galaxy group is at the center between component A and B, the component A and B are probably gravitationally lensed QSOs. We estimate that the Einstein mass is 1.46 $\times$ $10^{11}$ $M_{\odot}$ and the total mass of the lens is 1.34 $\times$ $10^{13}$ $M_{\odot}$. The deflection angle is 1.97 arc seconds at positions A and B and the velocity dispersion is 261\,$km\,s^{-1}$. Theoretically, this candidate could be a pair of fold images of a strong lensing system by a galaxy group, and we will investigate the possibility when the redshifts of nearby galaxies are available.
 }

\authorrunning{Y. H. Chen, M. Y. Tang, H. Shu, H. Tu}       
\titlerunning{A possible gravitational lensing phenomenon}  
\maketitle

\section{Introduction}

Gravitationally lensed QSOs open a window for extragalactic and cosmological phenomena. The strongly lensed QSOs systems can be used as an important tool to study the mass distribution of dark matter in galaxies or galaxy clusters (Gray et al. 2000, Richard et al. 2010, and their follow up papers). The geometric structure of gravitationally lensed QSOs can be used to study the sources (Ding et al. 2017) and the lensing galaxies (Oguri, Rusu \& Falco 2014). The general theory of relativity shows that the distortion of space and time near some massive celestial body causes the deflection of light when passing through the massive celestial body. The gravitationally lensed QSOs present multiply-lensed QSOs or Einstein rings. Lemon et al. (2017) performed a research of resolving small-separation gravitationally lensed QSOs in Global Astrometric Interferometer for Astrophysics (GAIA). In GAIA DR2, Lemon et al. (2019) discovered 22 new gravitationally lensed QSOs. Wen et al. (2009) discovered 4 gravitational lensing systems in the Sloan Digital Sky Survey (SDSS) data release 6 (DR6). Inada et al. (2006) identified SDSS J1029+2623 as a gravitationally lensed QSO with an image separation of 22.5 arc seconds. Misawa et al. (2016) studied the narrow absorption lines in the lensed QSO SDSS J1029+2623 by multi-sightline observation. Oguri \& Marshall (2010) reported the potential of wide-field optical imaging surveys to search for gravitationally lensed QSOs and supernovae. Lapi et al. (2012) derived analytical formulae based on model calculations of galaxy-scale gravitational lensing. The gravitational lensing phenomena are very valuable astronomical phenomena.

The Large Sky Area Multi-Object Fiber Spectroscopic Telescope (LAMOST), also named Guo Shou Jing Telescope (GSJT), is a wide-field of view ($\sim$5 degrees) and a large effective aperture ($\sim$4 meters) optical telescope (Cui et al. 2012). It is an innovative active reflecting Schmidt telescope developed by Chinese scientists and operated by National Astronomical Observatories, Chinese Academy of Sciences. LAMOST spectral survey focuses on the experiment for galactic understanding and exploration, and the extragalactic survey (Zhao et al. 2012). From 2011 to 2021, the LAMOST data release 9 (DR9) v1.0 has released spectra of 10,907,516 stars, 242,569 galaxies, and 76,167 QSOs, and so on. This provides rich candidates for the research work in various fields of astrophysics. Chen, Li, \& Shu (2022) performed a statistical research of global parameters for AFGK type stars released by LAMOST DR8 and compared them with the theoretical calculation models. Napolitano et al. (2020) performed a first research of central velocity dispersion of almost 86,000 galaxies in LAMOST DR7. From the 'Unknown' data set in LAMOST DR7, Lu et al. (2022) identified 106 new emission line galaxies and 29 new galactic H\,II regions. Chen et al. (2021) penetrated into the Mg\,II absorption line systems in QSO spectra from LAMOST DR1-5. Chen, Tang \& Shu (2023) derived an initial statistical analysis of LAMOST DR9 low resolution QSOs. Based on the LAMOST, a large number of achievements are produced every year.

There are 76,167 QSOs identified in the LAMOST DR9 v1.0. We plan to search for gravitationally lensed QSOs based on these QSOs initially. We select LAMOST J1606+2900 as a strongly lensed QSOs candidate. The photometric and spectra data of LAMOST J1606+2900 have been analyzed in detail. A quick lens modeling is implemented for LAMOST J1606+2900 and some basic parameters of lens galaxies have been initially studied.

This paper is organized as follows. In Sect. 2, we introduce a basic method of selecting candidates, including two parts of datasets and methodology. The basic information of a selected candidate QSO (LAMOST J1606+2900) is displayed in Sect. 3. Then, we give a discussion and conclusions in Sect. 4.

\section{Candidate selection}

\subsection{Datasets}
We plan to try to find gravitationally lensed QSOs based on the QSOs released by LAMOST DR9 v1.0. The LAMOST pipelines of 'v2.9.7' is used to process and analyze spectra for DR9 v1.0, which contains a spectra reduction pipeline 'v2.9' and a spectra analysis pipeline 'v7' (Luo et al. 2015, Ai et al. 2016). The spectra of QSOs are from LAMOST Low-Resolution Spectroscopic Survey with a resolution of 1,800 at 5,500 {\AA} (Stoughton et al. 2002, Abazajian et al. 2003). These spectra cover from 3,690 {\AA} to 9,100 {\AA}. Both LAMOST and SDSS carry on ugriz photometric observations and spectral sky surveys, so their data can complement each other. The data released by LAMOST includes the source\_ID and g-band magnitude values of the target source in GAIA, which can be used as supplementary tests. At the LAMOST website, there is a link named CDS portal, which contain the existing multi band photometric observations for the target source. Among them, the Dark Energy Camera (DECam) provides an efficient option for obtaining photometry from 4,000 {\AA} to 10,000 {\AA}. The Dark Energy Spectroscopic Instrument (DESI) is being conducted on the Mayall 4-meter telescope at Kitt Peak National Observatory. The DESI is designed to measure the expansion history of the universe over the past 10 billion years and to study the equation of state of dark energy and its time evolution (Dey et al. 2019).

\subsection{Methodology}

The coordinate values and redshift values can be used to select possible gravitationally lensed QSOs initially. First, we download the parameter data of 76,167 QSOs and delete the redshift values of poor quality spectra. If the coordinate values are too close, we will pick out the observations of the same QSO among different years. Therefore, we select the right ascension difference and declination difference of QSOs in a range of 1.44 to 14.4 arc seconds respectively as an initial attempt. At the same time, we constrain the redshift difference of the selected QSOs to be less than 0.1. There are 23 sets of target sources were screened out. Then, the name, brightness, spectrum, photometry and other information of each QSO will be visually checked carefully. Special attention should be paid to check whether there are groups of galaxies, gravitationally lensed arcs, Einstein crosses, or Einstein rings near the QSOs. We eliminate the target sources with only one optical counterpart and there are 8 sets of target sources left. We eliminate 5 sets of target sources with the differences of g-band magnitude in GAIA data close to or greater than 1.0. There are 3 sets of target sources left, LAMOST J112501.37+325622.7 and LAMOST J112501.89+325619.8, LAMOST J160603.01+290050.8 and LAMOST J160602.81+290048.7, and LAMOST J121641.76+292529.2 and LAMOST J121642.24+292537.9. For the first set of target sources, no galaxies were found within the range of 5 arc minutes in the simbad astronomical database. For the third set of target sources, the redshift value is 2.5867 and 2.5000, being 3.5\% different. This method is very initial. We need to find out that there are galaxies around the QSOs, and the magnitude values of Gaia g-band for the QSOs are close, as well as that of SDSS ugriz-bands. At last, LAMOST J160603.01+290050.8 (A) and LAMOST J160602.81+290048.7 (B) are selected as a candidate and we named it as LAMOST J1606+2900. In other sky survey observations, a higher proportion of QSO pairs were obtained using more complex methods, such as Lemon et al. (2023) for Gaia, Valeri \& Nathan (2023) for Gaia and Wise, and Eftekharzadeh et al. (2017) for SDSS. Our method is simple and initial, and we will improve our search method in future work.

Myers et al. (2008) identified LAMOST J1606+2900 as quasar pair based on SDSS DR4. Inada et al. (2010) performed follow-up observations in i-band with Astrophysical Research Consortium (ARC) 3.5 m telescope and identified them as binary quasars because no lens galaxy between the two QSO components was found. However, we find a galaxy group near LAMOST J1606+2900. If the dark matter center of the galaxy group is just at the center of component A and B, the dark matter can make LAMOST J1606+2900 a gravitationally lensed QSOs. The detailed information is presented in Sect. 3.

\section{The basic information of a selected candidate LAMOST J1606+2900}

\begin{table*}
\caption{Properties of LAMOST J1606+2900 (component A and B) and the nearby galaxies and stars.}
\begin{center}
\begin{tabular}{lllllllllll}
\hline
Celestial body                &R.A.        &Decl.      &Gaia$\_$g    &u        &g        &r        &i        &z        &Redshift                   \\
\hline
QSO(A)                        &16 06 03.01 &29 00 50.8 &18.413652    &18.64    &18.39    &18.32    &18.39    &18.29    &0.7670                     \\
                              &            &           &             &         &         &         &         &         &$\pm$0.0002                \\
                              &            &           &             &         &         &         &         &         &                           \\
QSO(B)                        &16 06 02.81 &29 00 48.7 &18.171162    &18.83    &18.48    &18.41    &18.33    &18.27    &0.7684                     \\
                              &            &           &             &$\pm$0.02&$\pm$0.01&$\pm$0.01&$\pm$0.01&$\pm$0.03&$\pm$0.0099                \\
                              &            &           &             &         &         &         &         &         &                           \\
$\frac{ABS(A-B)}{A}$          &            &           &1.3$\%$      &1.0$\%$  &0.5$\%$  &0.5$\%$  &0.3$\%$  &0.1$\%$  &0.2$\%$                    \\
                              &            &           &             &         &         &         &         &         &                           \\
\hline
Galaxy1                       &16 06 02.39 &29 01 08.4 &             &22.99    &23.07    &21.94    &21.74    &23.14    &                           \\
                              &            &           &             &$\pm$0.83&$\pm$0.28&$\pm$0.17&$\pm$0.21&$\pm$1.38&                           \\
Galaxy2                       &16 06 02.56 &29 01 12.4 &             &23.86    &22.52    &21.83    &21.71    &20.52    &                           \\
                              &            &           &             &$\pm$1.22&$\pm$0.14&$\pm$0.11&$\pm$0.15&$\pm$0.27&                           \\
Galaxy3                       &16 06 04.40 &29 00 57.5 &             &20.50    &18.77    &17.94    &17.57    &17.27    &                           \\
                              &            &           &             &$\pm$0.11&$\pm$0.01&$\pm$0.01&$\pm$0.01&$\pm$0.02&                           \\
Galaxy4                       &16 06 05.03 &29 00 45.6 &             &25.07    &23.87    &23.48    &21.93    &20.91    &                           \\
                              &            &           &             &$\pm$1.26&$\pm$0.36&$\pm$0.39&$\pm$0.16&$\pm$0.32&                           \\
Galaxy5                       &16 06 06.15 &29 00 55.0 &             &23.27    &22.84    &20.94    &20.20    &19.86    &                           \\
                              &            &           &             &$\pm$0.75&$\pm$0.17&$\pm$0.05&$\pm$0.04&$\pm$0.14&                           \\
Galaxy6                       &16 06 05.64 &29 00 36.4 &             &23.74    &23.54    &22.61    &22.15    &21.66    &                           \\
                              &            &           &             &$\pm$0.91&$\pm$0.26&$\pm$0.17&$\pm$0.17&$\pm$0.52&                           \\
Galaxy7                       &16 06 03.20 &29 00 07.2 &             &23.42    &22.62    &21.32    &20.53    &19.87    &                           \\
                              &            &           &             &$\pm$1.04&$\pm$0.17&$\pm$0.08&$\pm$0.06&$\pm$0.17&                           \\
Galaxy8                       &16 06 00.95 &29 00 36.6 &             &22.81    &21.96    &20.74    &20.17    &19.73    &                           \\
                              &            &           &             &$\pm$0.53&$\pm$0.08&$\pm$0.04&$\pm$0.04&$\pm$0.12&                           \\
Galaxy9(SDSS                  &16 06 09.90 &28 58 03.4 &             &19.93    &18.48    &17.73    &17.29    &17.01    &0.09416                    \\
J160609.90+285803.4)          &            &           &             &$\pm$0.09&$\pm$0.01&$\pm$0.01&$\pm$0.01&$\pm$0.03&$\pm$0.00002               \\
Galaxy10(SDSS                 &16 06 17.66 &28 58 23.4 &             &19.33    &17.22    &16.21    &15.76    &15.41    &0.09204                    \\
J160617.66+285823.4)          &            &           &             &$\pm$0.06&$\pm$0.01&$\pm$0.00&$\pm$0.00&$\pm$0.01&$\pm$0.00002               \\
Galaxy11(SDSS                 &16 06 09.90 &29 04 17.2 &             &19.74    &19.47    &19.11    &18.98    &18.76    &0.25617                    \\
J160609.90+290417.2)          &            &           &             &$\pm$0.04&$\pm$0.01&$\pm$0.01&$\pm$0.01&$\pm$0.04&$\pm$0.00004               \\
\hline
Star1                         &16 06 02.67 &29 00 12.3 &             &25.01    &24.03    &22.63    &21.94    &21.79    &                           \\
                              &            &           &             &$\pm$1.06&$\pm$0.34&$\pm$0.16&$\pm$0.13&$\pm$0.51&                           \\
Star2                         &16 06 01.85 &29 00 10.1 &             &24.44    &23.43    &22.47    &22.16    &22.08    &                           \\
                              &            &           &             &$\pm$1.17&$\pm$0.23&$\pm$0.15&$\pm$0.17&$\pm$0.65&                           \\
Star3                         &16 06 02.87 &29 01 04.6 &             &22.87    &21.16    &19.70    &18.80    &18.32    &                           \\
                              &            &           &             &$\pm$0.43&$\pm$0.03&$\pm$0.02&$\pm$0.01&$\pm$0.03&                           \\
Star4                         &16 06 00.46 &29 01 11.3 &             &22.85    &20.17    &18.74    &18.00    &17.51    &                           \\
                              &            &           &             &$\pm$0.44&$\pm$0.02&$\pm$0.01&$\pm$0.01&$\pm$0.02&                           \\
Star5                         &16 06 00.39 &29 01 05.2 &             &24.02    &23.22    &22.89    &22.47    &21.81    &                           \\
                              &            &           &             &$\pm$0.97&$\pm$0.18&$\pm$0.20&$\pm$0.20&$\pm$0.51&                           \\
\hline
\end{tabular}
\end{center}
\end{table*}

In Table 1, we show the properties of LAMOST J1606+2900 (component A and B) and the nearby galaxies and stars. According to the right ascension and declination of component A and B of LAMOST J1606+2900, we can derive that the image separation is 3.36 arc seconds using the Python software package "astropy.coordinates". The gaia$\_$g magnitude of component B is only 1.3\% brighter than that of component A. The ugriz magnitudes of component A are from LAMOST and that of component B are from SDSS. For component A, there is no spectra from SDSS. For component B, there is only V magnitude of 18.72 from LAMOST. The ugriz magnitudes of component B differ from that of component A by less than or equal to 1.0\%, as shown in Table 1. For component A, the redshift is 0.7670$\pm$0.0002 from LAMOST. For component B, the redshift is 0.7686$\pm$0.0099 from LAMOST and 0.7682$\pm$0.0002 from SDSS. We take an average redshift of 0.7684$\pm$0.0099 for component B in Table 1, which is only 0.2\% larger than that of component A.

\begin{figure*}
\begin{center}
\includegraphics[width=9.5cm,angle=0]{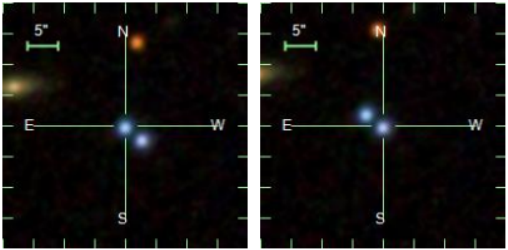}
\end{center}
\caption{A photometric image of LAMOST J160603.01+290050.8 (A, left) and LAMOST J160602.81+290048.7 (B, right) from SDSS.}
\end{figure*}

\begin{figure*}
\begin{center}
\includegraphics[width=9.5cm,angle=0]{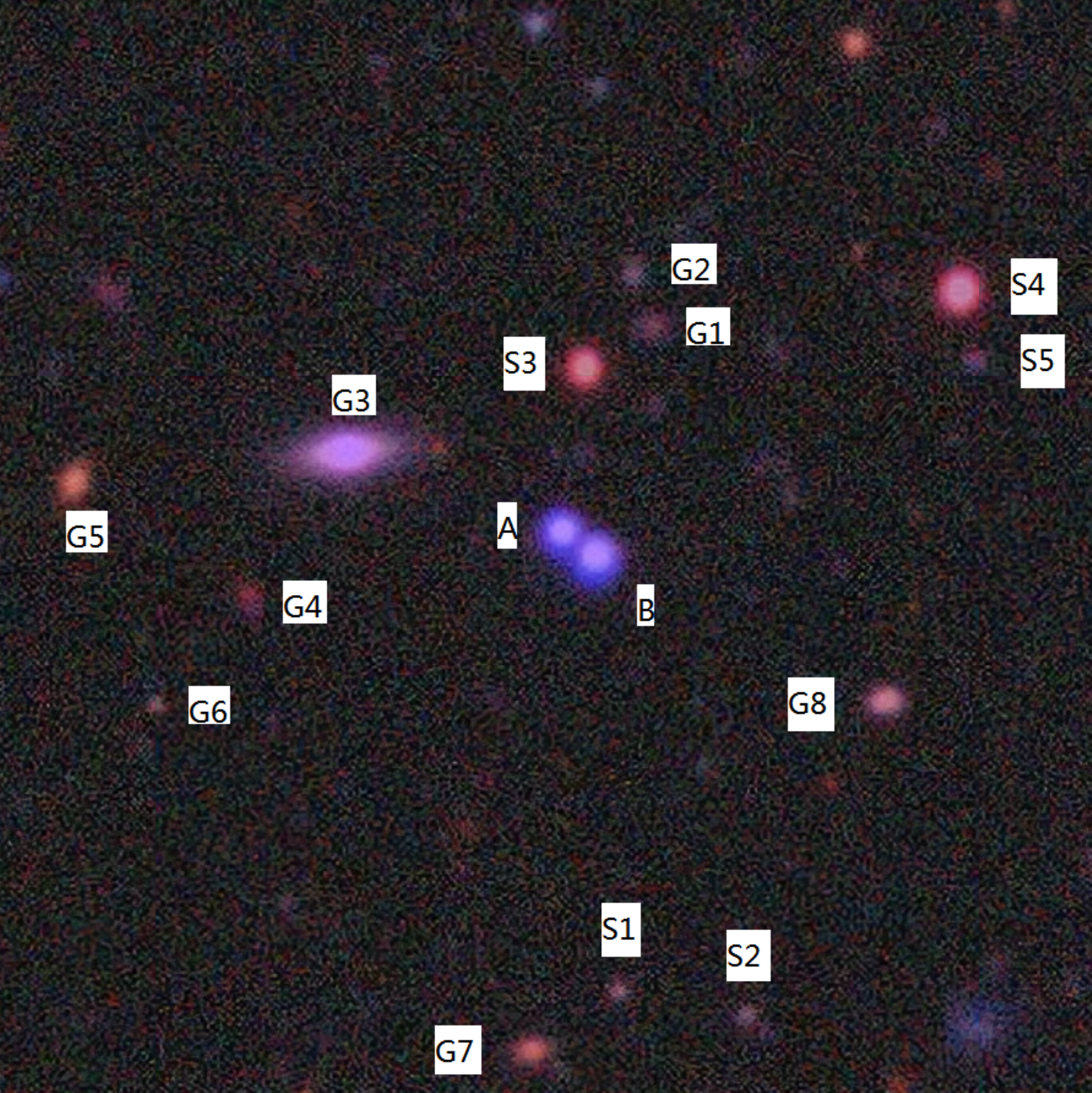}
\end{center}
\caption{A native color map for LAMOST J1606+2900 from the DECam Legacy Survey DR10. For short numbers, G1-8 represent galaxies and S1-5 are stars.}
\end{figure*}

In Fig. 1, we show a photometric image of LAMOST J160603.01+290050.8 (A, left) and LAMOST J160602.81+290048.7 (B, right) from SDSS. The image separation is 3.36 arc seconds for component A and B and both the component A and B are blue. However, we can not see an obvious optical target source between component A and B. Based on this, Inada et al. (2010) identified them as binary quasars.

In the CDS portal, we carefully examine the existing multi band photometric observations for LAMOST J1606+2900. In the DECam Legacy Survey (DECaLS DR5), we can obtain clearer photometry pictures. The latest DECaLS data release is DR10 and we cut a native color map around 1.69 arc minutes from DECaLS DR10 named as Fig. 2. The signal to noise ratio (SNR) is 476 for component B, 263 for component A, and only 42 between component A and B from DECaLS DR10. The SNR between component A and B is generated by the influence of component A and B and there is truly no optical target source between component A and B. In Fig. 2, we accidentally find that there are some galaxies around the LAMOST J1606+2900. We check the rainbow color map and check the celestial bodies from SDSS one by one. In Fig. 2, we use G1-8 to represent galaxies and S1-5 to represent stars. The coordinate values and magnitude values (ugriz) of these celestial bodies are shown in Table 1, which are from SDSS. Obviously, there is a galaxy group with some foreground stars. There is a galaxy group, but we don't have a spectra of these galaxies. The spectra of component A and B for LAMOST J1606+2900 should be studied deeply.

\begin{figure*}
\begin{center}
\includegraphics[width=11.5cm,angle=0]{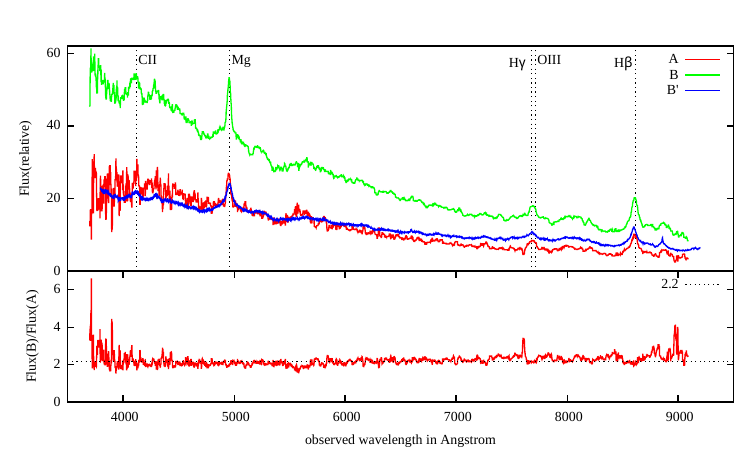}
\end{center}
\caption{A figure of relative flux of LAMOST J1606+2900. The spectra of A and B are from LAMOST and the spectra of B' are from SDSS.}
\end{figure*}

\begin{table*}
\caption{Emission lines and corresponding redshift values of component A and B from LAMOST. The emission line wavelengthes and spectra line center wavelengthes are from the LAMOST low resolution catalog DR9 v1.0 website.}
\begin{center}
\begin{tabular}{lllllllllll}
\hline
                                          &\multicolumn{2}{c}{Component A}    &\multicolumn{2}{c}{Component B}    \\
\hline
\hline
Emission lines                            &Line center     &Redshift          &Line center     &Redshift          \\
\hline
CII ($\lambda$\,2326\,{\AA})              &4111\,{\AA}     &0.7674            &4111\,{\AA}     &0.7674            \\
Mg ($\lambda$\,2800\,{\AA})               &4947\,{\AA}     &0.7668            &4951\,{\AA}     &0.7682            \\
H$\gamma$($\lambda$\,4341.6803\,{\AA})    &7668.3179\,{\AA}&0.76620971        &7682.4565\,{\AA}&0.76946619        \\
OIII($\lambda$\,4364.3782\,{\AA})         &7710.8120\,{\AA}&0.76676073        &7717.9170\,{\AA}&0.76838868        \\
H$\beta$($\lambda$\,4862.6778\,{\AA})     &8590.1377\,{\AA}&0.76654470        &8602.0137\,{\AA}&0.76898698        \\
\hline
\end{tabular}
\end{center}
\end{table*}

\begin{figure*}
\begin{center}
\includegraphics[width=9.5cm,angle=0]{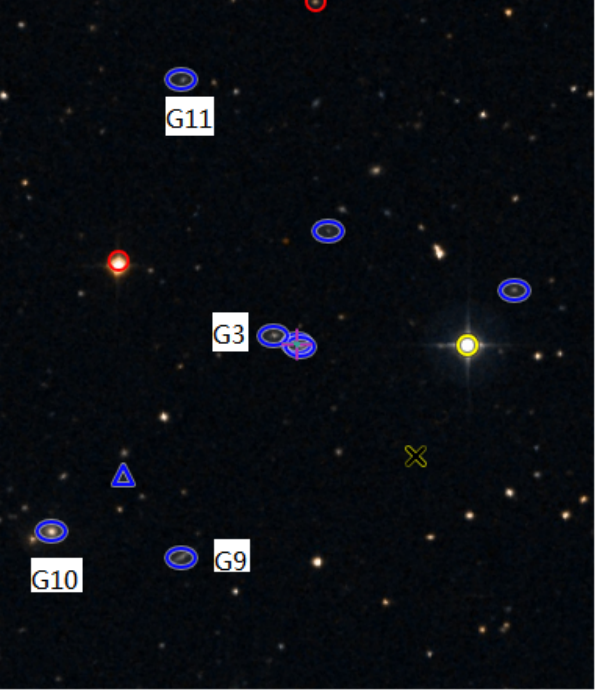}
\end{center}
\caption{A native color map for LAMOST J1606+2900 from the simbad astronomical database. The field of view is 7.73 arc minutes.}
\end{figure*}

We show a figure of relative flux of LAMOST J1606+2900 in Fig. 3. The abscissa is the observed wavelength in Angstrom covering from 3,690 {\AA} to 9,100 {\AA} and the ordinate is the relative flux. In the upper panel, the spectra of A and B are from LAMOST, while the spectra of B' are from SDSS. The emission lines of C\,II, Mg, H\,$\gamma$, O\,III, and H\,$\beta$ are identified among the spectra of A, B, and B'. These emission lines are very similar. In particular, H\,$\gamma$ line and O\,III line are partially overlapped in the upper panel. All the emission lines correspond to a redshift value of 0.767 for component A and 0.768 for component B, as shown in Table 2. No emission lines with other redshift values (possibly related to the galaxy group) are found. These emission lines are probably derived from the intrinsic emission related to LAMOST J1606+2900. In the lower panel, we show a ratio of flux(B) to flux(A) from LAMOST. It is basically a constant in the lower panel, around 2.2, which indicates that the emission lines and continuous spectrum of component B are proportionally amplified for that of component A. The component A and B are probably two gravitationally lensed images of the same quasar source. Inada et al. (2010) reported the i-band observations and the similar shapes of Mg II emission lines at z = 0.770 for LAMOST J1606+2900. Inada et al. (2010) identified them as binary quasars because there is no galaxy between component A and B. LAMOST shows the spectra from 3,690 {\AA} to 9,100 {\AA} for both component A and B of LAMOST J1606+2900. The emission lines of C\,II, Mg, H\,$\gamma$, O\,III, and H\,$\beta$ are present at z $\sim$ 0.767. No emission lines with other redshift values are found and no optical sources between component A and B are found.

There is a galaxy group around LAMOST J1606+2900. If the dark matter center of the galaxy group is just at the center of component A and B, then the component A and B are likely gravitationally lensed QSOs. There is no redshift values for G1-8 in Table 1. In the simbad astronomical database, we enlarge the field of view to 7.73 arc minutes and draw Fig. 4. In Fig. 4, the plus symbol in the center is the component A and B and the upper left of the center is G3. The closest galaxies with a redshift value are G9-11. The redshift value of G9-11 is 0.09416, 0.09204, and 0.25617 respectively from SDSS, as shown in Table 1. The G11 is a Seyfert 1 galaxy in the simbad astronomical database and we assume that the redshift values of the galaxy group are 0.09416. We assume the dark matter center of the galaxy group is just at the center of component A and B. 
The gravitational lensing phenomena fitting software "lenstool" is a widely used program for modeling group and cluster scale strong lensing systems (URL https://projets.lam.fr/projects/lenstool/wiki, Jullo et al. 2007). We try to fit the component A and B as gravitationally lensed QSOs with lenstool initially. The profile is adopted as circular singular isothermal sphere and elliptical singular isothermal sphere. We set $Z_{source}$ to 0.767 and $Z_{lens}$ to 0.09416. For cosmological constants, we assume that $H_{0}$ is 70\,km$\cdot$s$^{-1}$$\cdot$Mpc$^{-1}$, $\Omega_{m}$ is 0.3 and $\Omega_{\lambda}$ is 0.7. The "lenstool" does not fit giant arcs but multiple images because pixelized strong lens modeling for galaxy groups or clusters is unacceptably slow.

Although the fitting results are not optimistic, we can still carry out some fundamental analysis based on the fitting results. The angular diameter distance is 1526.02\,Mpc between the observer and the source (DOS), 1302.72\,Mpc between the lens and the source (DLS), and 360.61\,Mpc between the observer and the lens (DOL). For the galaxy group with $Z_{lens}$ = 0.09416, 1.0 arc second corresponds to 1.748 kpc. The Einstein radius ($\theta_{E}$) is 1.68 arc seconds and the Einstein mass is 1.46 $\times$ $10^{11}$ $M_{\odot}$. According to Eq.\,(45) ($\theta_{E} \times$ DOS = $\alpha \times$ DLS) of Ramesh \& Matthias (1996), the deflection angle ($\alpha$) is 1.97 arc seconds. Namely, the dark matter deflects the light by $\alpha$ = 1.97 arc seconds at positions A and B. The parameters are consistent with the equation of $\alpha = \frac{4GM}{c^{2}R}$ and therefore, they are self-consistent. In addition, we can derive the one-dimensional velocity dispersion $\sigma_{v}$ of 261\,$km\,s^{-1}$ according to Eq.\,(44) ($\alpha = 4 \pi \frac{\sigma_{v}^{2}}{c^{2}}$) of Ramesh \& Matthias (1996). The galaxy G10 is 242 arc seconds away from component A using the Python software package "astropy.coordinates". If we take the cutoff radius ($r_{cut}$) at position G10, the total mass of the lens is 1.34 $\times$ $10^{13}$ $M_{\odot}$ according to the Eq.\,(42) ($\frac{GM}{r_{cut}} = 2 \sigma_{v}^{2}$) of Ramesh \& Matthias (1996).

It seems unreasonable that we can not see any baryons in such a massive halo. The component A and B also might be a fold image-pair of the galaxy group depicted in Fig 2. With only two images (component A and B), it is too difficult to limit the fitting parameters of the strong gravitational lensing phenomenon of multiple galaxy mass clumps. If we try to fit, the fitting error will be very large and no more physical information will be provided. In addition, we lack the direct redshift values of galaxy G1-8. We will implement more detailed model fittings and investigate the possibility when the redshifts of nearby galaxies are available in the future.

\section{Discussion and conclusions}

In this paper, we download the QSO data of LAMOST DR9 low resolution catalog and try to search for gravitationally lensed QSOs. A total of 76,167 QSOs are downloaded and initially analyzed. As an initial attempt, we select the right ascension difference and declination difference of QSOs in a range of 1.44 to 14.4 arc seconds and we constrain the redshift difference of the selected QSOs to be less than 0.1 at the same time. Then, we visually check the name, brightness, spectrum, photometry and other information of each QSO selected. We eliminate the target sources with only one optical counterpart, the differences of g-band magnitude in GAIA data close to or greater than 1.0, and the differences of redshift values greater than 1.0\%. At last, the QSOs LAMOST J160603.01+290050.8 (component A) and LAMOST J160602.81+290048.7 (component B) are selected and named as LAMOST J1606+2900.

The component A and B are 3.36 arc seconds apart and they display blue in Fig. 1. The Gaia$\_$g, ugriz, and redshift values of component A and B are 1.3\%, less than or equal to 1.0\%, and 0.2\% consistent respectively, as shown in Table 1. For the spectra, the emission lines of C\,II, Mg, H\,$\gamma$, O\,III, and H\,$\beta$ are present for component A and B. In particular, H\,$\gamma$ line and O\,III line are partially overlapped, as shown in Fig. 3. The ratio of flux(B) to flux(A) from LAMOST is basically a constant, around 2.2. All the properties show that the component A and B seem to be two images of the same QSO. But there is no optical target source between component A and B. Based on this, the two QSOs are previously identified as binary quasars (Inada et al. 2010, Eftekharzadeh et al. 2017).

Inada et al. (2010) and Eftekharzadeh et al. (2017) identified LAMOST J160603.01+290050.8 and LAMOST J160602.81+290048.7 as binary quasars. By carefully examining the surrounding celestial bodies, we find that there are so many galaxies surrounded. We infer that these galaxies are a galaxy group. The two components are 3.36 arc seconds apart and they have a redshift value of $\sim$0.767. We use the redshift value of G9 ($Z_{lens}$ = 0.09416) to represent the redshift values of the galaxy group. We assume that the dark matter center of the galaxy group is just at the center of component A and B and the dark matter make one QSO source to be the component A and B with $Z_{source}$ = 0.767. By using the software lenstool, we try to fit the component A and B as gravitationally lensed QSOs and derive some fundamental analysis. The angular diameter distance is 1526.02\,Mpc between the observer and the source, 1302.72\,Mpc between the lens and the source, and 360.61\,Mpc between the observer and the lens. The Einstein radius is 1.68 arc seconds, the Einstein mass is 1.46 $\times$ $10^{11}$ $M_{\odot}$, and the deflection angle is 1.97 arc seconds at positions A and B. The one-dimensional velocity dispersion is 261\,$km\,s^{-1}$ and the total mass of the lens is 1.34 $\times$ $10^{13}$ $M_{\odot}$ when we take the cutoff radius at position G10.

The component A and B also might be a fold image-pair of the galaxy group depicted in Fig 2. But with only two images of component A and B, it is too difficult to limit the fitting parameters of the strong gravitational lensing phenomenon of multiple galaxy mass clumps.

We have not only found a candidate of strongly lensed QSOs from LAMOST but also analyzed it with available information and modeled it using Lenstool. In the future, we will improve our search method for gravitationally lensed QSOs, hoping to find more strongly lensed QSOs with more sufficient evidence. If the redshift values of G1-8 are obtained in some future sky survey projects, it will contribute to in-depth analysis of the system, especially if the redshift values of G1-8 remain close to that of G9, which will increase the credibility of our present analysis.

\section{Acknowledgment}

We are very grateful to an anonymous reviewer who strongly improved the original version of this work. Guoshoujing Telescope (the Large Sky Area Multi-Object Fiber Spectroscopic Telescope LAMOST) is a National Major Scientific Project built by the Chinese Academy of Sciences. Funding for the project has been provided by the National Development and Reform Commission. LAMOST is operated and managed by the National Astronomical Observatories, Chinese Academy of Sciences. We acknowledge the support of NSFC (11803004) and Yunnan Province Youth Talent Project (2019-182). M. Y. Tang acknowledges the support by NSFC through grants No.12203011, and Yunnan provincial Department of Science and Technology through grant No.202101BA070001-261.

\label{lastpage}
\end{document}